\documentclass[aps,prb,showpacs,superscriptaddress,onecolumn,preprint]{revtex4}%
\usepackage{amsfonts}
\usepackage{amsmath}
\usepackage{amssymb}
\usepackage{graphicx}%
\begin{document}
\title{Equation of state for shock compressed xenon in the ionization regime: ab
initio study}
\author{Cong Wang}
\affiliation{LCP, Institute of Applied Physics and Computational
Mathematics, P.O. Box 8009, Beijing 100088, People's Republic of
China}
\author{Yun-Jun Gu}
\affiliation{Laboratory for Shock Wave and Detonation Physics
Research, Institute of Fluid Physics, P.O. Box 919-102, Mianyang,
Sichuan, People¡¯s Republic of China}
\author{Qi-Feng Chen}
\affiliation{Laboratory for Shock Wave and Detonation Physics
Research, Institute of Fluid Physics, P.O. Box 919-102, Mianyang,
Sichuan, People¡¯s Republic of China}

\author{Xian-Tu He}
\affiliation{LCP, Institute of Applied Physics and Computational
Mathematics, P.O. Box 8009, Beijing 100088, People's Republic of
China} \affiliation{Center for Applied Physics and Technology,
Peking University, Beijing 100871, People's Republic of China}
\author{Ping Zhang}
\thanks {Corresponding author; zhang\_ping@iapcm.ac.cn}
\affiliation{LCP, Institute of Applied Physics and Computational
Mathematics, P.O. Box 8009, Beijing 100088, People's Republic of
China} \affiliation{Center for Applied Physics and Technology,
Peking University, Beijing 100871, People's Republic of China}

\begin{abstract}
Quantum molecular dynamic (QMD) simulations have been applied to
study the thermophysical properties of liquid xenon under dynamic
compressions. The equation of state (EOS) obtained from QMD
calculations are corrected according to Saha equation, and
contributions from atomic ionization, which are of predominance in
determining the EOS at high temperature and pressure, are
considered. For the pressures below 160 GPa, the necessity in
accounting for the atomic ionization has been demonstrated by the
Hugoniot curve, which shows excellent agreement with previous
experimental measurements, and three levels of ionization have
been proved to be sufficient at this stage.

\end{abstract}

\maketitle

\section{INTRODUCTION}

\label{sec-introduction}

Accurate understandings of the thermodynamic properties of materials under
extreme conditions are of great scientific interest \cite{PBX:Ernstorfer:2009}%
. The relative high temperatures (several eV) and densities (several
g/cm$^{3}$) produce the so-called \textquotedblleft warm dense
matter\textquotedblright\ (WDM), which provides an active research platform by
combining the traditional plasma physics and condensed matter physics. WDM
usually appears in shock or laser heated solids, inertial confinement fusion,
and giant planetary interiors, where simultaneous dissociations, ionizations,
and degenerations make modelling of the dynamical, electrical, and optical
properties of WDM extremely challenging.

The study of xenon in the warm dense region has long been focused because of
the potential applications in power engineering and the significance in
astrophysics \cite{PBX:Drake:2006}. Various experimental measurements
\cite{PBX:Keeler:1965,PBX:Nellis:1982,PBX:Fortov:1976,PBX:Gryaznov:1980,PBX:Urlin:1991a,PBX:Urlin:1991b,PBX:Eremets:2000}
and theoretical models
\cite{PBX:Radousky:1988,PBX:Kuhlbrodt:2005,PBX:Ross:1980,PBX:Schwarz:2005,PBX:Chen:2009}
have been applied to probe the EOS of xenon. Since the ionization equilibrium
are not interfered with the dissociation equilibrium between molecules and
atoms, xenon, which is characterized by monoatomic molecule and close shell
electronic structure, is particularly suitable for the investigation of the
high pressure behavior under extreme conditions. Despite its simple atomic
structure, modelling the EOS of xenon under dynamic compression, especially
the Hugoniot curve, is an essential issue. For instance, 0-K isotherm, which
is important in determining the physical properties of solid xenon under
static pressure, has been modelled by augmented-plane-wave (APW) electron band
theory method \cite{PBX:Ross:1980}. While, the EOS of xenon under dynamic
compression has been calculated through the fluid perturbation theory (FPT),
and compared with experimental data \cite{PBX:Ross:1980}. On the other hand,
the chemical picture of fluid xenon can be simply described through the
concept of plasma physics, deep insight into the shock compressed xenon has
been extracted by the self-consistent fluid variational theory (SFVT)
\cite{PBX:Chen:2009}.

However, the electronic structure has been convinced to be the key
role in determining the thermophysical properties of WMD
\cite{PBX:Wang:2010a}. Due to the intrinsic approximations of
these classical methods, fully quantum mechanical description of
xenon under shock compressions still remains to be presented and
understood. Meanwhile, QMD simulations, where quantum effects are
considered by the combinations of classical molecular dynamics for
the ions and density functional theory (DFT) for electrons, have
already been proved to be successful in describing WDM
\cite{PBX:Kietzmann:2008,PBX:Lorenzen:2009}.

In the present paper, QMD simulations, where free electrons
induced by the high temperatures and pressures are considered,
have been used to calculate the thermophysical properties of warm
dense xenon. The EOS are determined for a wide range of densities
and temperatures, and compared with experimental measurements and
different theoretical models. The Hugoniot curves, where different
levels of ionization are considered, are then derived from the
EOS. The rest of the paper is organized as follows. The simulation
details are briefly described in Sec. \ref{sec-method};
Corrections to QMD data and the EOS are presented in Sec.
\ref{sec-analysis}. Finally, we close our paper with a summary of
our main results.

\section{COMPUTATIONAL METHOD}

\label{sec-method}

In this work, we have performed simulations for shock compressed xenon by
employing the Vienna Ab-initio Simulation Package (VASP), which was developed
at the Technical University of Vienna \cite{PBX:Kresse:1993,PBX:Kresse:1996}.
$N$ atoms in a fixed volume supercell, repeated throughout the space
periodically, form the elements of simulations. After introducing
Born-Oppenheimer approximation, the thermo-activated electrons are fully
quantum mechanically treated through plane-wave, finite-temperature DFT
\cite{PBX:Lenosky:2000}. The electronic states are populated according to the
Fermi-Dirac distribution at temperature $T_{e}$. The exchange correlation
functional is determined by generalized gradient approximation (GGA) with the
parametrization of Perdew-Wang 91 \cite{PBX:Perdew:1991}. Projector augmented
wave (PAW) pseudopotential \cite{PBX:Blochl:1994} are used to present the
ion-electron interactions. The electronic structure are obtained from a series
of given ionic positions, which are subsequently varied according to the
forces calculated within the framework of DFT via the Hellmann-Feynman theorem
at each molecular dynamics step. The simulations are executed with the
isokinetic ensemble (NVT), where the ionic temperature ($T_{i}$) is controlled
by Nos\'{e} thermostat \cite{PBX:Nose:1984}, and the system is kept in local
thermo-dynamical equilibrium by setting the electron temperature $T_{e}$ equal
to $T_{i}$.

The bulk of the QMD simulations are performed using $\Gamma$ point to sample
the Brillouin zone in molecular dynamic simulations with the plane-wave cutoff
of 600.0 eV, because EOS (pressure and energy) can only be modified within 5\%
for the selection of higher number of \textbf{k} points and larger cutoff
energy. A number of 64 atoms is included in the simulation box. To cover
typical Hugoniot points with respect to the data obtained from experiments,
the densities adopted in our simulations range from 2.965 to 10.0 g/cm$^{3}$
and temperatures between 165 and 50000 K. Each density and temperature point
is typically simulated for 4 $\sim$ 6 ps, and the time step selected for the
integrations of atomic motion lies between 0.5 and 2.0 fs with respect to
different conditions. Then, the subsequent 1 ps at equilibrium are used to
calculate EOS as running averages.

\section{RESULTS AND DISCUSSION}

\label{sec-analysis}

\subsection{EFFECTS OF FREE ELECTRONS IN WARM DENSE XENON}

\label{sec-correction}

\begin{figure}[ptb]
\centering
\includegraphics[width=9.0cm]{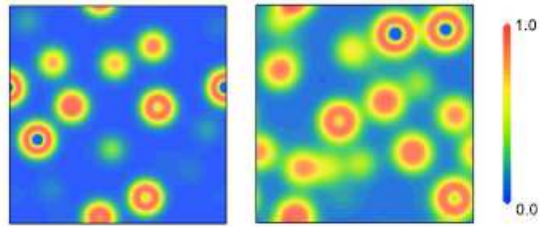}\caption{(Color online) Charge density
distribution for different conditions: $\rho$ = 5.0 g/cm$^{3}$, T = 1000 K
(left panel); and $\rho$ = 9.5 g/cm$^{3}$, T = 40000 K (right panel).}%
\label{fig_charge}%
\end{figure}

Internal energy for shock compressed materials can be expressed as:
\begin{eqnarray} \label{E}
E_{tot}=E_{ion}+T_{ion}+\int drV_{ext}(r)n(r)+\frac{1}{2}\int
drV_{H}(r)n(r)+E_{xc}+T_{electron}+\triangle E_{I},
\end{eqnarray}
where $E_{ion}$ and $T_{ion}$ correspond to the interaction energy
and the kinetic energy of the bare nuclear. The third term comes
from the fixed external potential $V_{ext}(r)$ (in most cases the
potential is due to the classical nuclei), in which the electrons
move. The fourth term is the classical electrostatic energy of the
electronic density, which can be obtained from the Hartree
potential. $E_{xc}$ and $T_{electron}$ denote exchange-correlation
energy and the kinetic energy of electrons. Finally, the ionization
energy is added in the last term, which stands for the energy to
generate free electrons and should be seriously considered in the
warm dense region. In QMD simulations, the last term is excluded,
and the simulated results should be corrected. The total internal
energy is then expressed as:
\[
E_{tot}=E_{QMD}+N\sum_{i}\alpha_{i}E_{i},
\]
where $N$ is the total number of atoms for the present system, and $\alpha
_{i}$ is the $i$-th level ionization degree, $E_{i}$ is the energy required to
remove $i$ electrons from a neutral atom, creating an $i$-level ion. Here,
$E_{i}$ equals to $\sum_{i}\vartriangle\epsilon_{i}$, and $\vartriangle
\epsilon_{i}$ is the $i$-th ionization energy.

Free electrons can be introduced at high temperatures and pressures
by subsequent ionization of atoms. The ionization precess can be
revealed from the change of the charge density distribution with
increasing density and temperature, as shown in Fig.
\ref{fig_charge}. In lower density and temperature region, the
localization of electrons represents the atomic fluid. While,
metallic behaviors of fluid xenon are indicated by delocalization of
electrons, which can be attributed to the thermo-activation of
electronic states. Similar phenomena have already been found for
another nobel gas helium \cite{PBX:Kietzmann:2007} and the expanded
fluid alkali metal Li \cite{PBX:Kietzmann:2008}.

\begin{figure}[ptb]
\centering
\includegraphics[width=9.0cm]{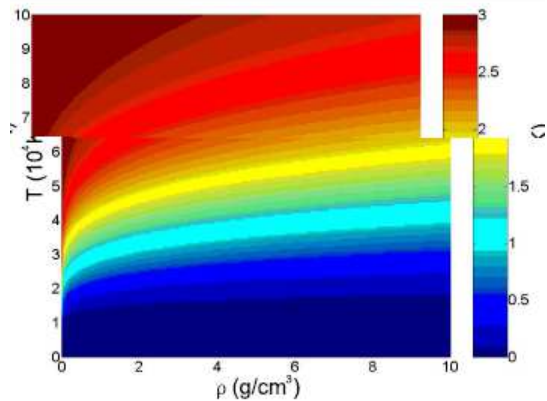}\caption{(Color online) The ionized
electron number as a function of the density and temperature.}%
\label{fig_ne}%
\end{figure}

Although ionizations of electrons can be qualitative described in
the framework of finite-temperature DFT, contributions to pressure
from noninteracting electrons are still lacking. Thus, the total
pressure should be corrected as:
\begin{align}
\label{tot_P}P_{tot}=P_{QMD}+\sum_{i}i\alpha_{i}\frac{\rho
k_{B}T}{m},
\end{align}
where $m$ presents the mass of xenon atom. The density and temperature are
denoted by $\rho$ and $T$ respectively, and $k_{B}$ stands for Boltzmann
constant. $E_{QMD}$ and $P_{QMD}$ are obtained from VASP. Since, it is
difficult to quantify the ionization degree for $i$-th level from direct QMD
results, as an alternative selection, we use Saha equation to verify the
degree of each ionization level:
\begin{equation}
\label{Saha}\frac{n_{i+1}n_{e}}{n_{i}}=\frac{2\Omega}{\lambda^{3}}%
\frac{g_{i+1}}{g_{i}}\exp(-\frac{\vartriangle\epsilon_{i}}{k_{B}T}),
\end{equation}
\begin{equation}
\label{lamda}\lambda=\sqrt{\frac{h^{2}}{2\pi m_{e}k_{B}T}},
\end{equation}
where $n_{i}$ is the density of atoms in the $i$-th state of
ionization, that is with $i$ electrons removed, $g_{i}$ is the
degeneracy of states for the $i$-ions, $n_{e}$ is the electron
density, $\lambda$ is the thermal de Broglie wavelength of an
electron, $m_{e}$ is the electron mass, and $\Omega$ is the volume
of the supercell. Here, $\alpha_{i}$ equals to $n_{i}\Omega/N$.
Similar approximations have already been used to study the EOS of
liquid deuterium under shock compressions \cite{PBX:Wang:2010b},
where soften behavior of the Hugoniot curve has been found at high
pressure, and the simulated results are consist with available
experiments.

\begin{figure}[ptb]
\centering
\includegraphics[width=9.0cm]{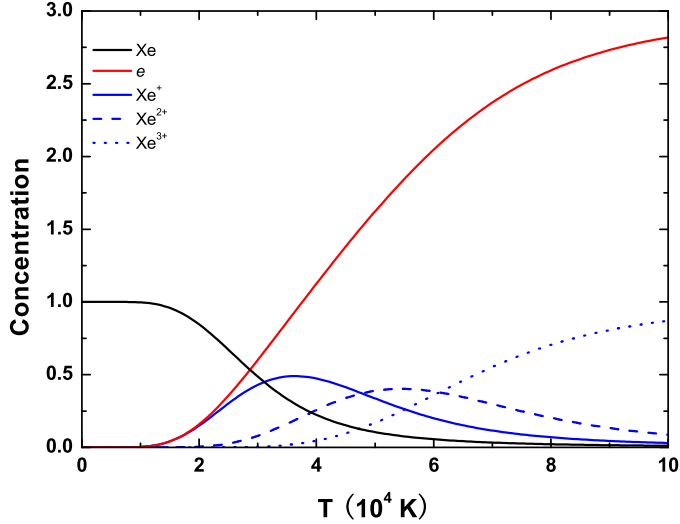}\caption{(Color online) Component of
plasma xenon as a function of temperature at $\rho$ = 6.0 g/cm$^{3}$.}%
\label{fig_component}%
\end{figure}

In the present work, three levels of ionization have been considered (the
relative ionization energies are 12.1, 21.2, and 32.1 eV respectively), and
the relative electron number (equals to $\sum_{i}i\alpha_{i}$) as a function
of density and temperature are labelled in Fig. \ref{fig_ne}. Xenon under
extreme conditions is considered to be partially ionized plasma, where
ionization equilibrium: Xe$^{(i-1)+}$ $\leftrightarrows$ Xe$^{i+}$+$e$, stands
for the changes in the electronic structure under high temperatures and
pressures. Three levels of atomic ionization have been taken into account in
the present calculations. As an example, the composition for density of 6.0
g/cm$^{3}$ is plotted in Fig. \ref{fig_component}. The fraction of charged
ions Xe$^{+}$, and Xe$^{2+}$ increase to maximum, then decrease, and the
number of Xe$^{3+}$ and electrons increase monotonously with temperature,
while the number of neutral atom Xe decrease with the increase of temperature.
The Hugoniot of shocked compressed liquid xenon can be effectively modified
with the consideration of atomic ionizations, and the detailed discussion can
be found in Sec. \ref{sec-eos}.

\subsection{THE EQUATION OF STATE}

\label{sec-eos}

\begin{figure}[ptb]
\centering
\includegraphics[width=9.0cm]{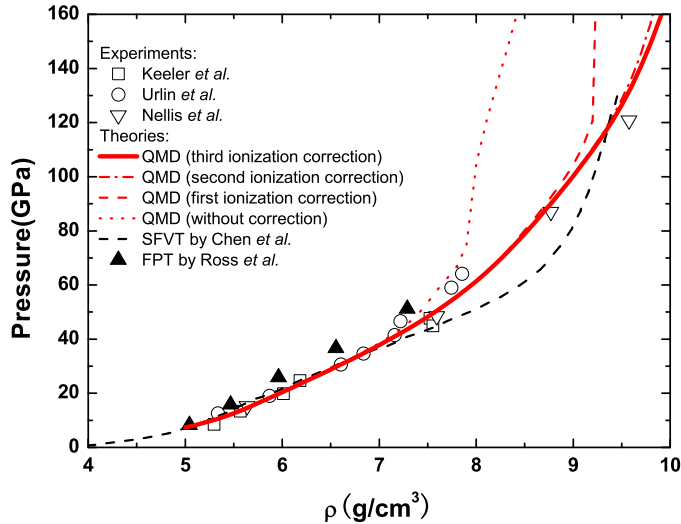}\caption{(Color online) Hugoniot for
shocked liquid xenon are compared with previous experimental
\cite{PBX:Keeler:1965,PBX:Nellis:1982,PBX:Urlin:1991a,PBX:Urlin:1991b} and
theoretical results \cite{PBX:Ross:1980,PBX:Chen:2009}. Specially, the
results, which are corrected according to different ionization level, are also
provided.}%
\label{fig_hugoniot}%
\end{figure}

Let us turn now to see the Hugoniot by investigating the corrected EOS of warm
dense xenon. Following Lenosky \emph{et al.}, Beule \emph{et al.}, and Holst
\emph{et al.} \cite{PBX:Lenosky:1997,PBX:Beule:1999,PBX:Holst:2008}, smooth
functions are used to fit the internal energies and pressures by expansions in
terms of density (g/cm$^{3}$) and temperature (10$^{3}$ K). The corrected QMD
data for internal energy (eV/atom) can be expanded as follows:
\begin{equation}
\label{E_expansion}E=\sum_{i=0}^{2}A_{i}(T)\rho^{i},
\end{equation}
\begin{equation}
\label{E_index}A_{i}(T)=a_{i0}\exp[-(\frac{T-a_{i1}}{a_{i2}})^{2}%
]+a_{i3}+a_{i4}T.
\end{equation}
The total pressure given in GPa can be similarly expanded as:
\begin{equation}
\label{P_expansion}P=\sum_{j=0}^{2}B_{j}(T)\rho^{j},
\end{equation}
\begin{equation}
\label{E_index}B_{j}(T)=b_{j0}\exp[-(\frac{T-b_{j1}}{b_{j2}})^{2}%
]+b_{j3}+b_{j4}T.
\end{equation}
The expansion coefficients $a_{ik}$ and $b_{jk}$ for $E$ and $P$
(accuracy better than 5\%) are summarized in Tab.
\ref{coefficient_E} and Tab. \ref{coefficient_P}, respectively.

\begin{table}[tbh]
\caption{Coefficients $a_{ik}$ in expansion for the internal energy $E$.}%
\label{coefficient_E}%
\centering
\begin{tabular}
[c]{cccccc}\hline\hline
$i$ & $a_{i0}$ & $a_{i1}$ & $a_{i2}$ & $a_{i3}$ & $a_{i4}$\\\hline
0 & 57.3976 & -17.3142 & 28.2460 & -38.0864 & 1.7336\\
1 & 11.2512 & 13.0002 & -67.2259 & -11.4532 & 0.0057\\
2 & -0.7180 & 41.0407 & 73.9168 & 0.6057 & 0.0060\\\hline\hline
\end{tabular}
\end{table}

\begin{table}[tbh]
\caption{Coefficients $b_{ik}$ in expansion for the total pressure $P$.}%
\label{coefficient_P}%
\centering
\begin{tabular}
[c]{cccccc}\hline\hline
$j$ & $b_{j0}$ & $b_{j1}$ & $b_{j2}$ & $b_{j3}$ & $b_{j4}$\\\hline
0 & -38.9090 & -15.8020 & 72.3514 & 63.7065 & -0.0607\\
1 & 0.3299 & 26.0136 & -5.7763 & -12.5064 & -0.0294\\
2 & 0.7302 & 1.0722 & 19.2058 & 0.9256 & 0.0441\\\hline\hline
\end{tabular}
\end{table}

A crucial measurement of the EOS data of xenon under shock conditions is the
Hugoniot, which can be derived from the following equation:
\begin{equation}
\label{hugoniot}(E_{0}-E_{1})+\frac{1}{2}(V_{0}-V_{1})(P_{0}+P_{1})=0,
\end{equation}
where $E$ is the internal energy, $P$ is the pressure, $V$ is the volume, and
the subscripts 0 and 1 denote the initial and shocked state, respectively. The
Hugoniot relation, which describes the locus of points in ($E, P, V$)-space,
follows from conservation of mass, momentum, and energy for an isolated system
compressed by a pusher at a constant velocity. In our simulations, the initial
density $\rho_{0}$ is 2.965 g/cm$^{3}$, the relative internal energy $E_{0}%
$=0.01 eV/atom at $T_{0}$=165 K. The initial pressure $P_{0}$ of the start
point on the Hugoniot can be approximated to be zero compared to high pressure
of shocked states.

\begin{figure}[ptb]
\centering
\includegraphics[width=9.0cm]{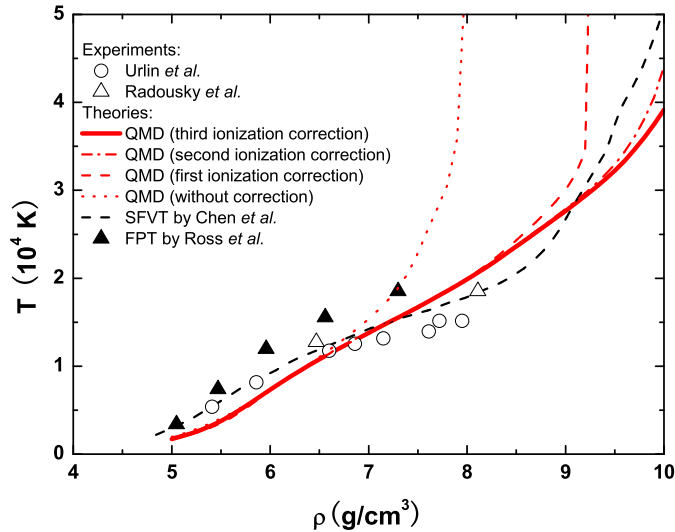}\caption{(Color online) Temperature is
labelled as a function of density along the Hugoniot. Previous experimental
\cite{PBX:Radousky:1988,PBX:Urlin:1991b} and theoretical
\cite{PBX:Ross:1980,PBX:Chen:2009} results are also included.}%
\label{fig_temperature}%
\end{figure}

The Hugoniot curve, where results from our simulations and
previous predictions from both experiments and theories are
labelled for comparison, is shown in Fig. \ref{fig_hugoniot}. In
order to clarify the influence of atomic ionization, results from
direct QMD simulations and those corrected by Saha equation are
also provided in the present work. Agreements can be found between
experimental measurements and the uncorrected QMD results only
below 70 GPa. However, at higher pressures ($P>$ 70 GPa), the
uncorrected Hugoniot curve shows an abrupt increase, which
suggests a stiff behavior, and unphysical results are obtained
here because of the ignorance of atomic ionization. Considerable
softening of the Hugoniot can be achieved by the corrected QMD
calculations, and reliable results are gradually revealed through
accounting for the first, second, and third level of ionization.
Here, the wide-range EOS for shock compressed liquid xenon can be
well described by the corrected QMD simulations (with the
consideration of three levels of ionization) below 160 GPa, where
good agreements are detected between our simulation results and
experiments
\cite{PBX:Keeler:1965,PBX:Nellis:1982,PBX:Urlin:1991a,PBX:Urlin:1991b}.
Theoretically, no Hugoniot points by previous FPT method
\cite{PBX:Ross:1980} are available to be used to compare with
experiments at pressures higher than 60 GPa, and the Hugoniot
curve from SFVT \cite{PBX:Chen:2009} seems to be inconsistent with
the results from Nellis \emph{et al.} \cite{PBX:Nellis:1982}.

Temperature, which is focused as one of the most important
parameters in experiments, is difficult to be measured because of
the uncertainty in determining the optical-intensity loss for
ultraviolet part of the spectrum in adiabatic or isentropic shock
compressions, especially for the temperature exceeding 20000 K.
QMD simulations provide powerful tools to predict shock
temperature. Our calculated Hugoniot temperature is shown in Fig.
\ref{fig_temperature} as a function of the density. For
comparison, the previous simulated (FPT \cite{PBX:Ross:1980} and
SFVT \cite{PBX:Chen:2009}) and experimental (by Radousky \emph{et
al.} \cite{PBX:Radousky:1988} and Urlin \emph{et al.}
\cite{PBX:Urlin:1991b}) results are also shown in Fig.
\ref{fig_temperature}. One can find that the present calculated
temperatures for the direct and corrected QMD simulations begin to
depart at about 10000 K (the respective density is 6.5
g/cm$^{3}$), which highlights the start point of ionization on the
Hugoniot. Discontinuous change in temperature, which is similar to
the behavior of Hugoniot, has been found in the direct QMD
simulation results, while, smooth variations of temperature along
the Hugoniot are predicted by the corrected simulation results.
The corrected shock temperatures are accordant with the available
experiments.

\section{CONCLUSION}

\label{sec-conclusion}

In summary, QMD simulations, where contributions from ionized
electrons are taken into account, are introduced to investigate the
EOS of shock compressed liquid xenon, and ionization equilibrium:
Xe$^{(i-1)+}$ $\leftrightarrows$ Xe$^{i+}$+$e$ is described by Saha
equation. Here, three levels of ionization are considered, and it
has been demonstrated that the accuracy of the calculated EOS can be
effectively improved through the present approximations. The
corrected EOS data are fitted by smooth functions, from which
Hugoniot curve are then derived. Good agreements have been achieved
between our calculated results and those from experimental
measurements.

\begin{acknowledgments}
This work was partially supported by the National High-Tech ICF
Committee of China.
\end{acknowledgments}


\end{document}